\documentclass[conference]{IEEEtran}
\usepackage{cite}
\usepackage{amsmath,amssymb,amsfonts}
\usepackage{algorithmic}
\usepackage{graphicx}
\usepackage{textcomp}
\usepackage{xcolor}
\usepackage[hyphens]{url}

\usepackage{xspace}
\usepackage{fancyhdr}
\usepackage{caption}
\usepackage{subcaption}
\usepackage{csquotes}
\usepackage{verbatim}
\usepackage{paralist}
\usepackage{subfiles}
\usepackage{enumitem}
\usepackage{soul}
\usepackage{booktabs}
\usepackage{tabularray}
\usepackage{color}
\usepackage{multirow}
\usepackage{multicol}
\def\BibTeX{{\rm B\kern-.05em{\sc i\kern-.025em b}\kern-.08em
    T\kern-.1667em\lower.7ex\hbox{E}\kern-.125emX}}

\newcommand{\ocf}{$\text{O}_{\text{CF}}$\xspace}
\newcommand{\ecf}{$\text{E}_{\text{CF}}$\xspace}
\title{Photonics for Sustainable Computing}
\author{
\IEEEauthorblockN{Farbin Fayza\IEEEauthorrefmark{1}, 
Satyavolu Papa Rao\IEEEauthorrefmark{2},
Darius Bunandar\IEEEauthorrefmark{3},
Udit Gupta\IEEEauthorrefmark{4},
Ajay Joshi\IEEEauthorrefmark{1}\IEEEauthorrefmark{3}}

\IEEEauthorblockA{\IEEEauthorrefmark{1}Boston University}
\IEEEauthorblockA{\IEEEauthorrefmark{2}NY CREATES}
\IEEEauthorblockA{\IEEEauthorrefmark{3}Lightmatter}
\IEEEauthorblockA{\IEEEauthorrefmark{4}Cornell Tech}
}
\begin{document}
\maketitle
\thispagestyle{plain}
\pagestyle{plain}

%%%%%% -- PAPER CONTENT STARTS-- %%%%%%%%

\begin{abstract}
Photonic integrated circuits are finding use in a variety of applications including optical transceivers, LIDAR, bio-sensing, photonic quantum computing, and Machine Learning (ML).
In particular, with the exponentially increasing sizes of ML models, photonics-based accelerators are getting special attention as a sustainable solution because they can perform ML inferences with multiple orders of magnitude higher energy efficiency than CMOS-based accelerators.
However, recent studies have shown that hardware manufacturing and infrastructure contribute significantly to the carbon footprint of computing devices, even surpassing the emissions generated during their use.
For example, the manufacturing process accounts for 74\% of the total carbon emissions from Apple in 2019.
This prompts us to ask -- if we consider both the embodied (manufacturing) and operational carbon cost of photonics, is it indeed a viable avenue for a sustainable future?
So, in this paper, we build a carbon footprint model for photonic chips and investigate the sustainability of photonics-based accelerators by conducting a case study on ADEPT, a photonics-based accelerator for deep neural network inference.
Our analysis shows that photonics can reduce both operational and embodied carbon footprints with its high energy efficiency and at least 4$\times$ less fabrication carbon cost per unit area than 28 nm CMOS.
\end{abstract}

\section{Introduction}
% \begin{figure}[t]
%     \centering
%      \includegraphics[width=0.48\textwidth]{samples/figures/teaser_draft.png}
%      \vspace{-0.2in}
%     \caption{[Draft values] Silicon photonics vs. CMOS for different aspects. (Need to find a good caption)}
%     \vspace{-0.2in}
%     \label{fig:teaser}
% \end{figure}
Photonic integrated circuits have proliferated over the past decade, seeing use in optical transceivers, LIDAR, bio-sensing, photonic quantum computing, and machine learning.
In particular, in the machine learning domain, photonics enable multiple orders of magnitude better energy efficiency for Deep Neural Network (DNN) inference~\cite{shiflett2021albireo, demirkiran2023electro, peng2022deep}, which is pervasively used these days.
% With the pervasiveness of DNNs, there is 
% Silicon photonics have also been used as accelerators, say for matrix multiplication, and for neural networks.
% Deep Neural Networks (DNNs) are extensively applied in various complex domains such as natural language processing, autonomous driving, security, and many more.
% This widespread adoption of DNNs is driven by a substantial exponential growth in the number of parameters within DNNs, making them more effective and versatile in tackling complex tasks.
% As Moore's Law comes to an end, conventional CMOS-based computing devices struggle to scale effectively with the increasing complexity of the DNNs and provide a sustainable computing environment.
% The rising favor for silicon-photonics stems from its ability to address this challenge, offering energy efficiency that surpasses traditional CMOS by multiple orders of magnitude \missingcitation.
%As an example, Figure \ref{fig:teaser} compares the average energy efficiency of a photonics-based accelerator, ADEPT, with an equivalent CMOS-based accelerator, considering the full system including data movement and computing~\cite{demirkiran2023electro}.
%On average ADEPT provides \hl{??} better energy efficiency, which indicates that silicon-photonics-based computing can be a potential path for sustainable computing.
%This energy efficiency reduces the operational carbon cost at the same scale.
% Therefore, in this era of DNNs, silicon photonics-based computing fosters optimism toward a sustainable computing solution.
However, being energy efficient during its use is not enough to claim that a technology is sustainable.
Recent studies have shown that the total carbon footprint of a chip is significantly dominated by the carbon emission during the fabrication process (embodied carbon), surpassing the carbon emission during operation (operational carbon) \cite{gupta2022act}.
For example, $\approx$74\% of Apple's annual carbon footprint in 2019 originated from the manufacturing process \cite{gupta2022act}.
From iPhone 3 to iPhone 11, the operational carbon decreased by 2.5$\times$ due to the efforts in optimizing energy efficiency, but the increasing complexity in the fabrication process led to a rise in the total carbon footprint by $\approx$1.4$\times$ \cite{gupta2021chasing}.
Therefore, it is useful to embark on a comprehensive study of the carbon footprint of photonic chips, encompassing both operational and embodied carbon, to determine whether photonics as currently envisioned is truly a viable computing solution for DNNs.
% This study would play a critical role in guiding the development of silicon photonics toward lower embedded carbon scenarios.

Over the past few years, numerous studies have explored the carbon footprint of CMOS chips and CMOS-based systems, but there are no studies that examine the carbon footprint of photonic chips.
Therefore, we first develop a carbon footprint model for photonic chips.
This model will help us understand the environmental impact of photonics.
%that can be heterogeneously integrated with electronics for acheiving the required system functionality.
To build the model, we utilize the average carbon cost of various processing steps (from the IMEC IEDM 2020 paper\cite{bardon2020dtco}), coupled with our estimate of a process flow for fabricating a reasonably complex photonic chip. 
With the model, we estimate that the embodied carbon cost per unit area of fabrication for a photonic chip is 4.1$\times$ lower than a 28 nm CMOS chip.
%Next, we consider an example electro-photonic DNN accelerator, ADEPT~\cite{demirkiran2023electro}, where an analog photonic chip is heterogeneously integrated with a digital electronic chip (required to store weights and for input/output activations), and with off-chip laser sources. 
Next, we consider an example electro-photonic DNN accelerator, ADEPT~\cite{demirkiran2023electro}, where an analog photonic chip with off-chip laser sources is integrated with a digital electronic chip (required to store data and process inputs/outputs).
We evaluate ADEPT to determine if photonics has the potential for sustainability when considering both the operational and embodied carbon.
Our analysis shows that, for one million inferences of several popular DNN models, ADEPT achieves 2.19$\times$ lower carbon footprint on average, with 14.58\% lower embodied carbon than an equivalent CMOS-based accelerator, despite consuming more area due to having photonic components.
At a broader level, this promising initial result indicates that a more in-depth evaluation of the carbon footprint of photonic technology is warranted.

\section{Carbon Footprint of a Chip}
\begin{figure}[t]
    \centering
     \includegraphics[width=0.49\textwidth]{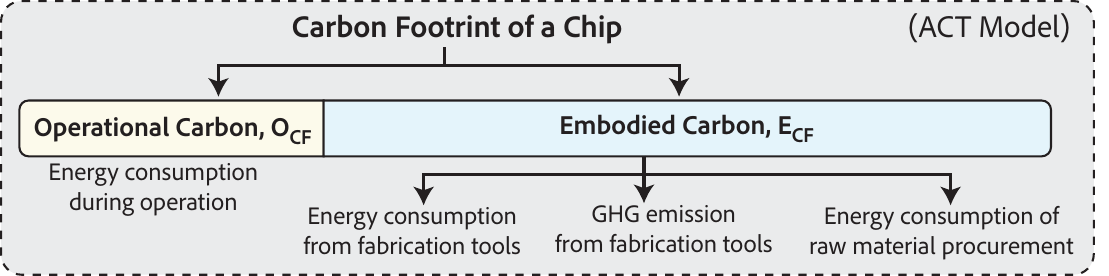}
    \vspace{-0.1in}
    \caption{Overview of the carbon footprint of a chip.}
    \vspace{-0.2in}
    \label{fig:cf-overvoew}
\end{figure}

In Figure \ref{fig:cf-overvoew}, we show an overview of the components that contribute to the carbon footprint of a chip following the well-established Architectural Carbon Modeling Tool (ACT)~\cite{gupta2022act}.
As mentioned before, the carbon footprint of a chip has two parts: operational carbon footprint (\ocf) and embodied carbon footprint (\ecf).
While the total carbon footprint is the summation of these two, we can amortize \ecf based on an application runtime and lifetime (as shown in Equation \ref{eqn:cf}) to estimate the carbon footprint of specific applications.

\vspace{-0.15in}
\small{
\begin{eqnarray}
    \text{CF} &=& \text{O}_{\text{CF}} +
\frac{\text{runtime}}{\text{lifetime}} \times \text{E}_{\text{CF}}\label{eqn:cf}\\
\text{O}_{\text{CF}} &=& \text{Energy}_{\text{use}} \times \text{CI}_{\text{use}}\label{eqn:ocf}\\
\text{E}_{\text{CF}} &=& \frac{\text{Area}}{\text{Yield}} \times (\text{EPA}\times\text{CI}_{\text{fab}} + \text{GPA} + \text{MPA}) + C_{\text{package}}\label{eqn:ecf}
\end{eqnarray}
}

\normalsize
\ocf is the product of the energy consumption for running an application with the carbon intensity (g/kW-hour) of the power source during the operation (CI$_{\text{use}}$) (Equation \ref{eqn:ocf}).
\ecf encompasses several factors including material extraction, energy consumption during fabrication, and associated greenhouse gas emissions (GHGs) (Equation \ref{eqn:ecf}).
Here, EPA is the energy consumption of the manufacturing tools per unit area of the chip, which is multiplied by the carbon intensity of the power source used during fabrication (CI$_{\text{fab}}$) to get the carbon cost of the fabrication process.
GPA is the GHG emission per unit area from some of the fabrication tools that use GHGs and release a portion of them into the air.
MPA comes from the cost of procurement of raw materials per unit area.
We add these three carbon components per unit area and multiply that with the effective chip area (considering yield).
Lastly, there is a packaging overhead carbon cost, C$_{\text{package}}$.
This is computed for the combination of chips that are heterogeneously integrated into one package; or for every chip, when they are placed separately on a PCB.
% Firstly, the energy consumption of the tools used in the manufacturing process contributes to the embodied carbon.
% Multiplying the energy consumption of the tools for fabricating a unit area (EPA) with the carbon intensity of the electricity source used in the fabrication process (CI$_{\text{fab}}$) gives us the fabrication carbon cost of the tools.
% Secondly, GPA, GreenHouse Gases (GHG) per unit ara of the chip fabrication, is caused by some of the manufacturing steps that use GHG and release a fraction of them in the air after use.
% Lastly, 

\section{Embodied Carbon of a Photonic Chip}
\begin{figure}[t]
    \centering
     \includegraphics[width=0.45\textwidth]{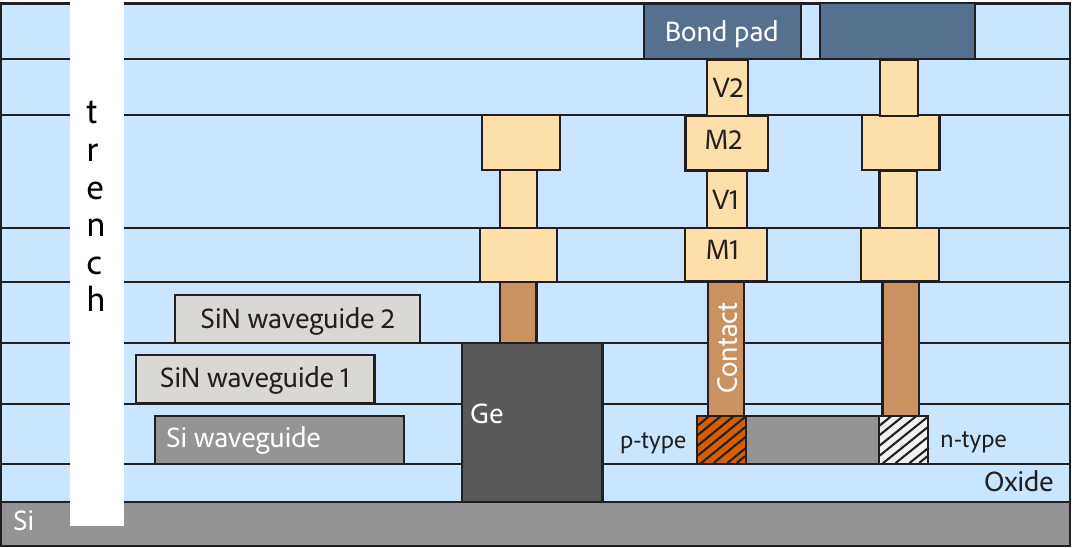}
    \caption{Cross-section of a photonic chip adapted from AIM Photonics multi-project wafer technology~\cite{fahrenkopf2019aim}.}
    \vspace{-0.15in}
    \label{fig:cross-section}
\end{figure}
While the operational carbon of a photonic chip can be estimated from the energy consumption and carbon intensity of the energy source, we need to analyze its fabrication process to calculate the embodied carbon.
To this end, we consider the cross-section of a photonic chip (see Figure \ref{fig:cross-section}) with off-chip laser sources. 
Here, we postulate a fabrication scheme that would result in a schematic cross-section that roughly matches that shown by Fahrenkopf \textit{et al.}~\cite{fahrenkopf2019aim}. 
The schematic cross-section is representative of a type of 'active' photonic circuit, due to the inclusion of modulators. 
%It also indicates how the complexity of an electronic chip can be avoided by using photonics where appropriate.

\subsection{Process Layers and Manufacturing Steps}
For a photonic chip fabricated using our example process, starting from a Silicon-on-Insulator (SOI) wafer, there are silicon (Si) and silicon-nitride (SiN) layers for fabricating the waveguides that facilitate optical signals. 
A germanium (Ge) photodiode layer is formed, typically by epitaxy, for sensing the processed light - the lower bandgap of germanium, when compared to silicon allows for the detection of light at a wavelength where silicon is transparent.
%To control the optical signals inside the Si waveguide we need to integrate a Mach-Zehner interferometer (MZI) that uses electro-optical methods.
%For this, in the MZI the p-type and n-type regions are formed in the silicon. 
To control the optical signals inside the Si waveguide (utilizing optical computing components such as Mach-Zehner interferometers or micro-ring resonators)  electro-optically, p-type and n-type regions are formed in the silicon.
The Ge photodiode similarly has doped regions (forming a p-i-n diode). 
These are connected to heterogeneously integrated electrical chips through metal layers (M1 and M2) and via layers (V1 and V2).
Bond pads are placed on top to allow the chip to be connected to others chips.\footnote{It should be noted that through-silicon vias would be needed for connection in a 3D system.}
Finally, a deep trench is created to expose the edge of the waveguide and allow the connection of optical fibers to the photonic waveguides at the edge of the photonic chip.

In general, for all the layers, the components are fabricated with the necessary lithography, etching, and cleaning steps, followed by polishing and oxide deposition (CVD).
The p-type and n-type regions are generated by a combination of implantation of dopant atoms into the silicon matrix (through openings in lithographically patterned photoresist) and thermal activation of the dopants. 
Openings (formed by reactive ion etch) on the wafer surface for metal and vias are filled with an ultra-thin copper diffusion barrier and copper using a combination of deposition and metal Chemical-Mechanical Planarization (CMP) processes. 
Additionally, the oxide that covers the waveguides is typically planarized by dielectric CMP. 
We collect the average energy consumption per wafer for each manufacturing step from the work by imec~\cite{bardon2020dtco}. 
It should be noted that differences in oxide thickness exist between typical values for photonic and electronic chips. 
Thicknesses used in photonic chips are similar to the highest metallization levels (10th to 16th metal layers in advanced chips) due to the need to avoid optical coupling between adjacent levels. 

\vspace{-0.05in}
\subsection{GHGs, Materials, and Packaging}
Several manufacturing steps such as dry etching and CVD chamber cleaning require GHGs with high global warming potential.
After utilizing the GHGs, the remaining gases pass through abatement systems, prior to being released into the atmosphere. 
%The degree of abatement can vary due to specific equipment utilized and adds some uncertainty to the GHG impact estimate for both electronic and photonic chips.
From ACT, we consider the GPA for a 28 nm technology node as the GPA of a photonic chip.
This is so that we can have a conservative estimate with a publicly available technology node with higher fabrication steps than photonics.
In actuality, it is likely to be far less because a 65 nm technology node (but using 193 nm optical lithography for resolving small spaces) can suffice for photonics.

Following ACT again, the MPA and packaging cost are independent of the technology node for a CMOS chip and we set their values to be the same as the ACT model for photonic chips.
Estimating the actual MPA and packaging cost for a photonic chip would require industry data and comprehensive analysis.

\subsection{Yield}\label{subsec:yield}
We use the Poisson model for calculating the yield of a photonic chip (as well as a CMOS chip that we use for comparison later in the paper).
Photonic circuits typically use wide spaces between waveguides (to avoid optical coupling), except in small portions of the circuit area associated with splitters, or points where coupling between elements is required, like a micro-ring resonator next to a waveguide.
Hence the nanometer size defects that are typical of advanced fabs do not have the same deleterious impact on photonic chips as on densely packed electronic chips.
Taking the splitter discussed by Zhang~\cite{zhang2019adjoint} as an example, the critical area can be estimated to be 20\% of the total chip area.
This critical area corresponds to the area occupied by the waveguides and the area around the splitters where Zhang \textit{et al.} point out the defect would have a significant impact.
Hence, the same defect density has a yield impact that is lower by a factor of $e^{-0.2}$ or 0.81 in a photonic chip than in an electronic chip of the same area, as shown in Figure \ref{fig:epa-bars} (right).
We use 0.1 $cm^{-2}$ defect density for our work.

\section{Photonics to Reduce Carbon}
\subsection{Benefits on Operational Carbon}

\renewcommand{\tabcolsep}{5pt}
\renewcommand{\arraystretch}{1}
\begin{table}[tb]
\caption{List of popular photonics-based DNN accelerators and their operational efficiency over CMOS-based systems.}
\vspace{-0.1in}
\label{tab:photonic-accelerators}
\begin{tabular}{@{}ll@{}}
\toprule
%\rowcolor[HTML]{9FC5E8} 
\textbf{Accelerator} & \textbf{Improvements over CMOS-based accelerators}                                                                                     \\ \midrule
ADEPT\cite{demirkiran2023electro} & \begin{tabular}[c]{@{}l@{}}Average 5.73$\times$ higher throughput/Watt than\\ systolic arrays considering a full system analysis\end{tabular} \\
\hline
Albireo\cite{shiflett2021albireo}              & \begin{tabular}[c]{@{}l@{}}Average 110$\times$ lower latency and 74.2$\times$ lower\\ EDP than CMOS-based accelerators\end{tabular}    \\
\hline
DEAP-CNN\cite{bangari2019digital}             & \begin{tabular}[c]{@{}l@{}}2.8 - 14$\times$ faster with $\approx$25\% less energy than GPUs\end{tabular}                                                                         \\
\hline
RecLight\cite{sunny2022reclight}             & \begin{tabular}[c]{@{}l@{}}37$\times$ lower energy-per-bit and 10\% better \\ throughput than CMOS-based RNN accelerators\end{tabular} \\
\hline
DNNARA\cite{peng2022deep}               & \begin{tabular}[c]{@{}l@{}}More than 19$\times$ faster than GPUs under \\ same power budget\end{tabular}                               \\ \bottomrule
\end{tabular}
\end{table}

%Photonics can efficiently perform several mathematical operations such as matrix-matrix multiplication, dot product, etc. that are the core building blocks of DNNs.
The key advantage for the energy efficiency of photonics over CMOS comes from the ability to compute in the optical domain at the speed of light with lower resistive and capacitive losses than in electrical devices.
%One of the reasons for higher energy efficiency in photonic computing than electronic computing is the lack of resistive and capacitive losses in photonic devices as against electrical devices.
% With efficient design of switching elements in photonics, energy efficiency also arises at the computational level. 
% The key advantage in energy efficiency of photonics over CMOS comes from the ability to compute in the optical domain at the speed of light with almost no power dissipation for computing.
% Additionally, by using light of different wavelengths to encode multiple inputs, photonic accelerators can perform parallel computation with high bandwidth.
Unfortunately, photonic systems are bottlenecked by their dependency on electric components.
Given there is no photonic memory yet, we need to convert data from electrical to photonic domain and vice versa for running a photonic computing device.
Also, the non-linear operations (ReLU, tanH, etc) of DNNs cannot be efficiently performed in photonics, so we need to switch to the electrical domain to perform them.
These conversions are costly, especially when we need to maintain the high bit precision requirements of DNN.
The dependency on electronic components also limits the throughput of the photonic systems, because the electronic components would be extremely power-consuming to keep up with the computation speed of photonics.
As a result, the current photonic accelerators run at 5-10 GHz at most.
Despite these constraints, photonics can gain multiple orders of magnitude higher energy efficiency than fully CMOS-based systems. 
In Table \ref{tab:photonic-accelerators}, we provide a short list of recent photonic DNN accelerators and how they compare with CMOS-based systems.
Researchers are actively working on different optimization strategies to overcome the above-mentioned challenges to design efficient photonic accelerators.

\subsection{Benefits on Embodied Carbon}
\begin{figure}[t]
    \centering
    \vspace{-0.1in}    \includegraphics[width=0.49\textwidth]{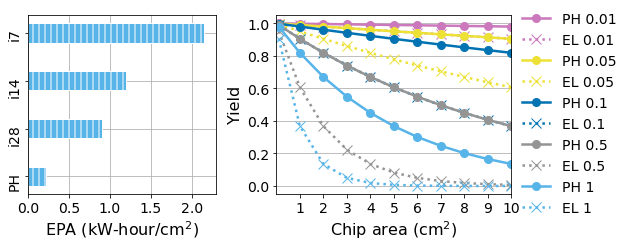}
     \vspace{-0.22in}
    \caption{Comparison of EPA (left) and yield for different defect densities ($cm^{-2}$) (right) of photonics (PH) and CMOS (EL) chips.}
    \vspace{-0.2in}
    \label{fig:epa-bars}
\end{figure}
%(Refer to Figure \ref{fig:epa-bars} for describing the items when applicable.)
% \begin{itemize}
%     \item Talk about photonics having less number of metal layers than CMOS
%     \item Follow by having less number of total fab steps
%     \item Talk about no need for EUV lithography
%     \item Talk about yield benefits
%     \item If possible also talk about lifetime.
% \end{itemize}
% \vspace{1.3in}

In Figure \ref{fig:epa-bars}, we show the energy consumption per unit area of fabrication of a photonic chip using our model, and a CMOS chip of 28 nm, 14 nm, and 7 nm technology node.
The photonic chip has fewer layers than CMOS chips overall, and so the fabrication complexity of a photonic chip is simpler than the recent CMOS chips. 
As a result, the EPA is 4.1$\times$, 5.6$\times$, and 9.8$\times$ less than a 28 nm, 14 nm, and 7 nm technology node, respectively.
Additionally, photonics does not require complex and power-consuming lithography technologies such as Extreme Ultra-Violet (EUV) lithography, which are required in the latest technology nodes (from 8 nm).
The yield is also higher for photonic chips because only 20\% of the total chip area is susceptible to defects as described in Section \ref{subsec:yield}.
Therefore, even though a photonic chip usually consumes more area than an equivalent CMOS chip, the embodied carbon cost is lower.

\section{A Case Study on ADEPT}

Demirkiran et al.~\cite{demirkiran2023electro}, present a comprehensive comparison between an electro-photonic accelerator, ADEPT, and electronic systolic arrays using power, performance, and area metrics on three popular DNN models: ResNet-50, BERT-Large, and RNN-T.
ADEPT uses a 128$\times$128 core at 10 GHz, while the equivalent systolic array design consists of 10 128$\times$128 systolic arrays running at 1 GHz.
We use the data from the ADEPT paper to calculate the carbon cost of one million inferences using our model for photonic components, and ACT for the electronic components.
All electronic components use a 22 nm technology node according to the paper.
We assume a lifetime of 5 years for both systems~\cite{lyu2023myths}.
The fabrication phase uses Taiwan's average power grid and the operation phase uses renewable power.

\begin{figure}[tb]
    \centering
     \includegraphics[width=0.48\textwidth]{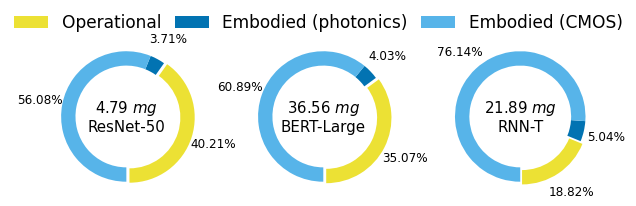}
   %  \vspace{-0.15in}
    \caption{Breakdown of ADEPT's carbon footprint for one million inferences (amortized embodied carbon).}
    %While the photonic components consume $\approx$16\% of the total area, they contribute to only $\approx$6\% of the embodied carbon.}
    \vspace{-0.2in}
    \label{fig:adept-cf-breakdown}
\end{figure}

We first show the breakdown of the carbon components of ADEPT in Figure \ref{fig:adept-cf-breakdown}.
While the embodied carbon is indeed dominating the carbon footprint, the contribution of the photonic embodied carbon is much lower than the electronic embodied carbon. 
This is because of the large electronic SRAMs used in the system as well as the low fabrication carbon cost and high yield of photonic chips.
Though photonic components consume $\approx$16\% of the total area of ADEPT, they contribute only $\approx$6\% to the embodied carbon with our pessimistic carbon model.
Note that the embodied-operational percentages may vary based on the power source \cite{gupta2022act}, but the message that we want to provide is that the embodied carbon does not become worse with the use of photonics.
Rather, replacing an electronic component with a photonic component in a system helps to reduce the embodied carbon, as well as the operational carbon.

\begin{figure}[tb]
    \centering
     \includegraphics[width=0.47\textwidth]{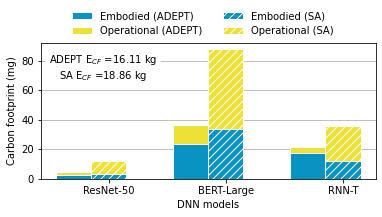}
   %  \vspace{-0.15in}
    \caption{Comparison of the carbon footprint of ADEPT (one 128$\times$128 core at 10 GHz) and equivalent Systolic-Arrays (SA) (10 128$\times$128 cores at 1 GHz) for one million inferences (amortized embodied carbon). SA's carbon footprint is 2.19$\times$ higher than ADEPT on average. Photonics wins in both operational and embodied carbon efficiency.}
    \vspace{-0.2in}
    \label{fig:adept-sa-compare}
\end{figure}

To understand how the carbon cost of an electro-photonic accelerator compares with an electronic accelerator, we compare the carbon footprint of ADEPT with equivalent systolic arrays (SA) in Figure \ref{fig:adept-sa-compare}.
On average, the carbon footprint of ADEPT is 2.19$\times$ lower than SAs.
%On SAs too, a significant portion of the area is taken by the SRAMs.
Even though ADEPT's area is $\approx$25 mm$^2$ larger than SAs, its embodied carbon footprint is 2.75 kg (14.58\%) lower because the photonic components reduce the embodied carbon with their simpler fabrication and higher yield than CMOS.
Moreover, SAs suffer from high operational carbon cost as well.
This is because running 10 CMOS-based cores to compete with the 10 GHz photonic core takes up a lot of power making the SAs energy-inefficient.
As a result, with both lower embodied carbon and lower operational carbon, ADEPT wins against SAs on the overall carbon footprint.


\section{Future Directions}
As DNNs become pervasive in various domains like autonomous driving, communication, and language processing, analyzing the carbon cost of the underlying hardware is critical to ensure that these technological advancements do not harm our environment.
In this study, we demonstrate the potential of photonics to offer a sustainable solution for DNNs, even under a conservative assessment.
The imminent crucial step is to have a comprehensive examination of photonic chip fabrication data across the various photonic manufacturing plants, enabling the development of a precise carbon footprint model for photonic chips.
The model should encompass all relevant factors, including fabrication processes, chiplet integration methods, GHG emissions, materials, and packaging.
An accurate carbon footprint model would allow computer architects to understand the design aspects of photonics that aid the environmental impact.
Considering the advantages and limitations of photonics, along with a carbon footprint model, various research directions emerge such as designing a carbon-aware balanced system with photonics and electronics, choosing proper design parameters and optimization metrics, choosing suitable workloads and platforms for CMOS and photonics, and many more.
Moving forward, as computer architects, we need to actively engage in research along these directions to make the best use of photonics for a sustainable environment.

%%%%%%% -- PAPER CONTENT ENDS -- %%%%%%%%

%%%%%%%%% -- BIB STYLE AND FILE -- %%%%%%%%
\bibliographystyle{IEEEtranS}
\bibliography{refs}
%%%%%%%%%%%%%%%%%%%%%%%%%%%%%%%%%%%%

\end{document}